\documentclass[12pt]{article}

\usepackage{scicite}

\usepackage[dvips]{graphicx}

\newenvironment{sciabstract}{%
\begin{quote} \bf}
{\end{quote}}

\newcounter{lastnote}

\title{Distinct Fermi-Momentum Dependent Energy Gaps in Deeply Underdoped Bi2212}

\author
{Kiyohisa Tanaka,$^{1,2}$ W.S.~Lee,$^{1}$ D.H.~Lu,$^{1}$ A.~Fujimori,$^{3}$\\
T.~Fujii,$^{4}$ Risdiana,$^{5}$ I.~Terasaki,$^{5}$\\
D.J.~Scalapino,$^{6}$ T.P.~Devereaux,$^{7,8}$ Z. Hussain,$^{2}$ and Z.-X.~Shen,$^{1\ast}$\\
\\
\normalsize{$^{1}$Department of Physics, Applied Physics, and Stanford Synchrotron Radiation Laboratory,}\\
\normalsize{Stanford University, Stanford, CA 94305, USA}\\
\normalsize{$^{2}$Advanced Light Source, Lawrence Berkeley National Lab, Berkeley, CA 94720, USA}\\
\normalsize{$^{3}$Department of Physics and Department of Complexity Science and Engineering,}\\
\normalsize{University of Tokyo, Kashiwa, Chiba 277-8561, Japan}\\
\normalsize{$^{4}$Cryogenic Center, University of Tokyo, Bunkyo-ku,
Tokyo 113-0032,
Japan}\\
\normalsize{$^{5}$Department of Applied Physics, Waseda University, Tokyo 169-8555, Japan}\\
\normalsize{$^{6}$Department of Physics, University of California, Santa Barbara, CA 93106-9530, USA}\\
\normalsize{$^{7}$Department of Physics, University of Waterloo, Ontario N2L3G1, Canada}\\
\normalsize{$^{8}$Pacific Institute for Theoretical Physics, University of British Columbia,}\\
\normalsize{Vancouver, British Columbia, V6T 1Z1, Canada}\\
\\
\normalsize{$^\ast$To whom correspondence should be addressed;
E-mail:  zxshen@stanford.edu.} }

\date{}

\begin{document}


\baselineskip24pt


\maketitle


\begin{sciabstract}
We use angle-resolved photoemission spectroscopy applied to deeply
underdoped cuprate superconductors Bi$_2$Sr$_2$(Ca,Y)Cu$_2$O$_8$
(Bi2212) to reveal the presence of two distinct energy gaps
exhibiting different doping dependence. One gap, associated with the
antinodal region where no coherent peak is observed, increases with
underdoping - a behavior known for more than a decade and considered
as the general gap behavior in the underdoped regime. The other gap,
associated with the near nodal regime where a coherent peak in the
spectrum can be observed, does not increase with less doping - a
behavior not observed in the single particle spectra before. We
propose a two-gap scenario in momentum space that is consistent with
other experiments and may contain important information on the
mechanism of high-$T_c$ superconductivity.
\end{sciabstract}


The pseudogap phase of underdoped cuprates has proven to be an
important region for discoveries and surprises in the field of
high-transition temperature ($T_c$) superconductors
\cite{Timusk:Pseudogap}. Early angle-resolved photoemission (ARPES)
\cite{Loeser:Bi2212:pseudogap} and electron tunneling experiments
from lightly underdoped samples \cite{Renner:STM_pseudogap}
suggested that the pseudogap has similar characteristics to the
superconducting gap below $T_c$, consistent with the idea that the
pseudogap is a precursor to the $d_{x^2-y^2}$ superconducting state
but lacks pair phase coherence. In this scenario, below $T_c$ where
the phase coherence of pairs is established, the pseudogap smoothly
evolves into the superconducting gap. There is only one energy scale
in the system and is associated with the magnitude of the gap at the
antinode. This antinodal gap was found via ARPES \cite{Campuzano},
thermal conductivity \cite{Mike:thermal_conductivity:SC_gap}, and
tunneling measurements \cite{zasadzinski:Tunnel_doping} to increase
as the doping was reduced from optimum. However, the energy gap
obtained by Andreev reflection
\cite{Deutscher:Andreev_reflection:Twogap}, penetration
depth\cite{Panagopoulos:penetration_depth:twogap}, and recent Ramman
experiments \cite{M.Opel:Raman:TwoGap, Tacon:Raman:Twogap} of
cuprates exhibits the opposite trend with doping suggesting a rather
different scenario from the one-gap picture.

We present ARPES data for deeply underdoped
Bi$_2$Sr$_2$(Ca,Y)Cu$_2$O$_8$ (Bi2212) crystals with $T_c$ values of
50 K, 40 K, and 30 K, finding evidence for the existence of two
distinct energy gaps in the single-particle spectral function ( See
``Supporting Online Materials'' for details about materials and the
ARPES measurements setup). One gap manifests itself as a spectral
weight suppression near the Brillouin zone boundary (antinodal
region); this antinodal gap becomes larger with decreased doping,
consistent with previous ARPES studies \cite{Campuzano}. The other
gap is resolved near the diagonal of the zone (nodal region) where a
quasi-particle peak can be observed. We find that the gap size of
this near nodal gap determined by empirical methods does not
increase by decreasing doping level. We attribute our result as
evidence for two distinct Fermi momentum dependent energy gaps, with
the gap near the nodal region corresponding to the superconducting
gap and the gap near the antinodal region to the pseudogap state.

In Fig. 1, we show energy distribution curves (EDCs) along the Fermi
surface (FS) for the $T_c=50$ K, 40 K, and 30 K samples. Compared to
previous ARPES studies on underdoped Bi2212 with a similar doping
level \cite{Harris}, our new data have a much improved quality; even
for the most underdoped sample ($T_c=30$ K), a clear quasi-particle
(or coherence) peak can still be observed near the nodal region
(Fig. 1B), which has not been seen previously in Bi2212 for such a
low doping. This improvement makes possible a more quantitative data
analysis. We first note that the lineshape of the spectra shows a
marked change along the FS. The sharp coherence peak gradually loses
spectral weight when moving from the nodal region toward the
antinodal region. In the antinodal region, the peak disappears and
only a broad hump in the spectrum located far away from the Fermi
level ($E_{\rm{F}}$) can be observed for the $T_c=30$ K and 40 K
samples. This behavior is consistent with previous studies of other
underdoped cuprates such as La$_{2-x}$Sr$_x$CuO$_4$ (LSCO) and
Ca$_{2-x}$Na$_x$CuO$_2$Cl$_2$ (Na-CCOC)
\cite{Yoshida_lightly,Kyle:NaCCOC:comp}. Here, we operationally
define the Fermi arc as the region where one can see a peak in the
superconducting state EDCs. Clearly, as demonstrated in Figs. 1B-G
and the inset of Fig. 1H, the length of the Fermi arc decreases as
the doping level of the samples decreases.

In Fig. 1H, we plot the leading-edge gap, defined as the energy
difference between the mid-point of the spectral leading edge and
that of the nodal spectrum within the Fermi arc region. We find that
the leading-edge gap is smaller for the lower $T_c$ samples
suggesting that the energy gap associated with this Fermi arc region
decreases with decreased doping. This doping dependence can be
directly observed in the raw spectrum. The EDCs in the intermediate
region (Fig. 1E) show that both the coherence peak position and the
leading-edge gap of the $T_c=50$ K sample is larger than the
$T_c=40$ K and 30 K samples. Similar observation can also be deduced
from Fig. 1, C and D, which are all still within the Fermi arc
region. This gap-reduction trend within the Fermi arc region is
opposite to the doping dependence of the energy gap in the antinodal
region reported in previous ARPES studies \cite{Campuzano}, and is
in conflict with the doping dependence of the energy gap inferred
from thermal conductivity \cite{Mike:thermal_conductivity:SC_gap}
and tunneling measurements \cite{zasadzinski:Tunnel_doping}.

In the antinodal region, the spectrum is characterized by a
suppression of the spectral weight over a region which one
associates with a pseudogap. To illustrate the difference in the
doping dependence of the pseudogap and the energy gap associated
with the Fermi arc, we have plot symmetrized EDCs
\cite{symmetrization} for the three samples for an intermediate
Fermi arc k-point (point 8), and an antinodal k-point (point 16)
(Fig. 2A, B). As indicated by the shaded areas, it is clear that the
EDCs of point 16 show a larger gap with more underdoping while the
EDCs of point 8 exhibit the opposite trend with doping. The inset in
Fig. 2B shows that there is essentially no change in the EDC of the
$T_c=30$ K sample between the superconducting ($T=10$ K, blue line)
and the normal state ($T=50$ K, red line) in the antinodal region.
It also suggests that the broadening of the peak with temperature
does not shift the leading-edge position downwards in these deeply
underdoped samples. For a more comprehensive view of the trend, the
peak positions of EDCs relative to the node at various locations
near the intermediate region (point 6, 8, and 10) and the antinodal
region (point 16) are plotted (Fig. 2C). Clearly, the doping
dependence of the energy gap along the Fermi arc and the antinodal
region are different. This behavior suggests that the energy gaps
associated with the Fermi arc region and antinodal region represent
two distinct energy gaps arising from different mechanisms. The
antinodal gap appears to be related to the pseudogap and not related
to superconductivity; whereas, the nodal gap more likely represents
the ``real'' superconducting gap because of the existence of a
coherence peak in the spectrum. The distinction between the two gaps
becomes smaller and harder to observe towards optimum doping
\cite{Timusk:Pseudogap,Loeser:Bi2212:pseudogap,Renner:STM_pseudogap,Campuzano}.
While an earlier experiment near optimal doping has seen this same
trend, the result was attributed to an anharmonic term in the
$d$-wave gap \cite{Mesot}. Uncovering the distinct spectral
lineshape in these deeply underdoped samples makes it possible to
attribute to distinct energy gaps.

As the energy gap in the antinodal region is primarily dominated by
the pseudogap, it is not straightforward to estimate the magnitude
of the maximum superconducting gap $\Delta_0$. Here, we exploit a
phenomenological method to estimate $\Delta_0$. The EDCs along the
Fermi arc are first divided by the Fermi-Dirac (FD) function at the
measurement temperature convoluted with the experimental resolution
(Fig. 3A). In this way, one can track the peak position of the
single-particle spectral function without the complications of FD
cut-off near $E_{\rm{F}}$. We then plot the peak position of these
spectra with respect to the $d_{x^2-y^2}$ function ($\left|\cos
k_x-\cos k_y\right|/2$), as illustrated in the inset of Fig. 3A. It
can be seen that the peak positions of these FD-function divided
spectra lay on a straight line for $k_{\rm{F}}$ values in the nodal
region, suggesting that the superconducting gap around the nodal
region is consistent with this $d_{x^2-y^2}$ form. We then estimate
$\Delta_0$ by extrapolating this straight line to the boundary where
$\left|\cos k_x-\cos k_y\right|/2=1$, as shown in the inset of Fig.
3A. Using this method, we extract $\Delta_0$ for samples with
various doping levels and summarize the result in Fig. 3B. The
values of $\Delta_0$ determined in this way increases as the doping
changes from $T_c=30$ K  to $T_c=50$ K, consistent with the behavior
of the peak positions shown in Fig. 2C. It reaches a maximum at a
doping level of approximately 0.1 and then remains at the same size
(or even slightly decreases) for doping levels up to the optimal
doping level. We note that similar behavior appears in results
obtained from Nernst effect measurements \cite{Nernst}. Thus, taken
at face value \cite {fitting}, the extrapolation of the nodal region
data along with the behavior of the peak position at the points 6,
8, and 10 suggest that the gap characteristic of the nodal region
has a distinct doping dependence from that of the antinodal gap.
Specifically, the nodal region gap has a $d$-wave momentum
dependence with an amplitude that remains relatively constant for a
range of doping below optimal doping and then decreases as the
system becomes severely underdoped.

Now, let us summarize our momentum-space picture of this two-gap
scenario. Beginning with the deeply underdoped Bi2212, there is a
small Fermi arc where a peak can be observed in the EDCs at the
Fermi crossing points $k_{\rm{F}}$. This arc, centered at the nodal
$k_{\rm{F}}$ point, then extends out along the FS increasing in
length as the doping is increased. Along this arc, we find evidence
for a $k$-dependent gap, consistent with a superconducting
$d_{x^2-y^2}$ gap with an increasing magnitude as the hole doping
increases. Another pseudogap which is much larger and decreases with
doping dominates the antinodal region. We believe that the smaller
nodal region gap is the true superconducting gap because it exhibits
a peak in the EDC. The pseudogap may arise from another mechanism
such as Umklapp scattering by the antiferromagnetic correlation or
competing states, such as stripes \cite{Stripe1,Stripe2}, polaronic
\cite{Kyle:NaCCOC:Polaron,Mishchenko:Polaron} behavior, or a
charge-density-wave \cite{Hanaguri:NaCCOC:CDW:STM,Kyle:NaCCOC:comp}.

This two-gap scenario is consistent with other experiments. First,
thermodynamic data suggest a distinct pseudogap and a
superconducting gap \cite{Loram}. Second, recent Raman studies
\cite{M.Opel:Raman:TwoGap, Tacon:Raman:Twogap} also suggest energy
gaps extracted from B$_{1g}$ (dominated by the antinodal region) and
B$_{2g}$ symmetry (dominated by the nodal region) have different
origins with opposite doping dependence. In addition, contradictive
results of the superconducting gap deduced from different
experimental tools can also be resolved within this momentum-space
two-gap picture. Andreev reflection
\cite{Deutscher:Andreev_reflection:Twogap} and penetration depth
measurement \cite{Panagopoulos:penetration_depth:twogap} clearly
indicate that the superconducting gap declines with more underdoping
while tunneling spectroscopy
\cite{Renner:STM_pseudogap,zasadzinski:Tunnel_doping} shows the
opposite trend with the doping. Such inconsistencies are difficult
to explain using a one-gap scenario, but are naturally compatible
with the two-gap scenario. We suggest that some measurements, such
as Andreev reflection and the penetration depth, are sensitive to
the superconducting condensate itself; thus, the superconducting gap
near the nodal region is probed. On the contrary, STM spectrum is
more sensitive to the antinodal gap because of larger phase space.
Thus, the doping dependence of the energy gap obtained from STM,
Andreev reflection and penetration depth is different because they
are sensitive to different gaps in the FS.

This two-gap scenario has two important implications which could be
very important for developing a microscopic theory of high $T_c$
superconductivity. First, the pseudogap near the antinodal region in
these deeply underdoped samples is unlikely a precursor state of the
superconducting state. Second, our data suggest that the weakened
superconductivity in the deeply underdoped region arises not only
from the loss of phase coherence \cite{emery:phase_pair_cross} due
to the decrease in the superfluid density but also a weakening of
the pair amplitude. In this case, a mechanism for the
superconducting gap reduction could be related to the shrinkage of
the coherent FS with less doping leading to a smaller phase space
for pairing.

\begin{figure}[p]
\begin{center}
\includegraphics[width=170mm]{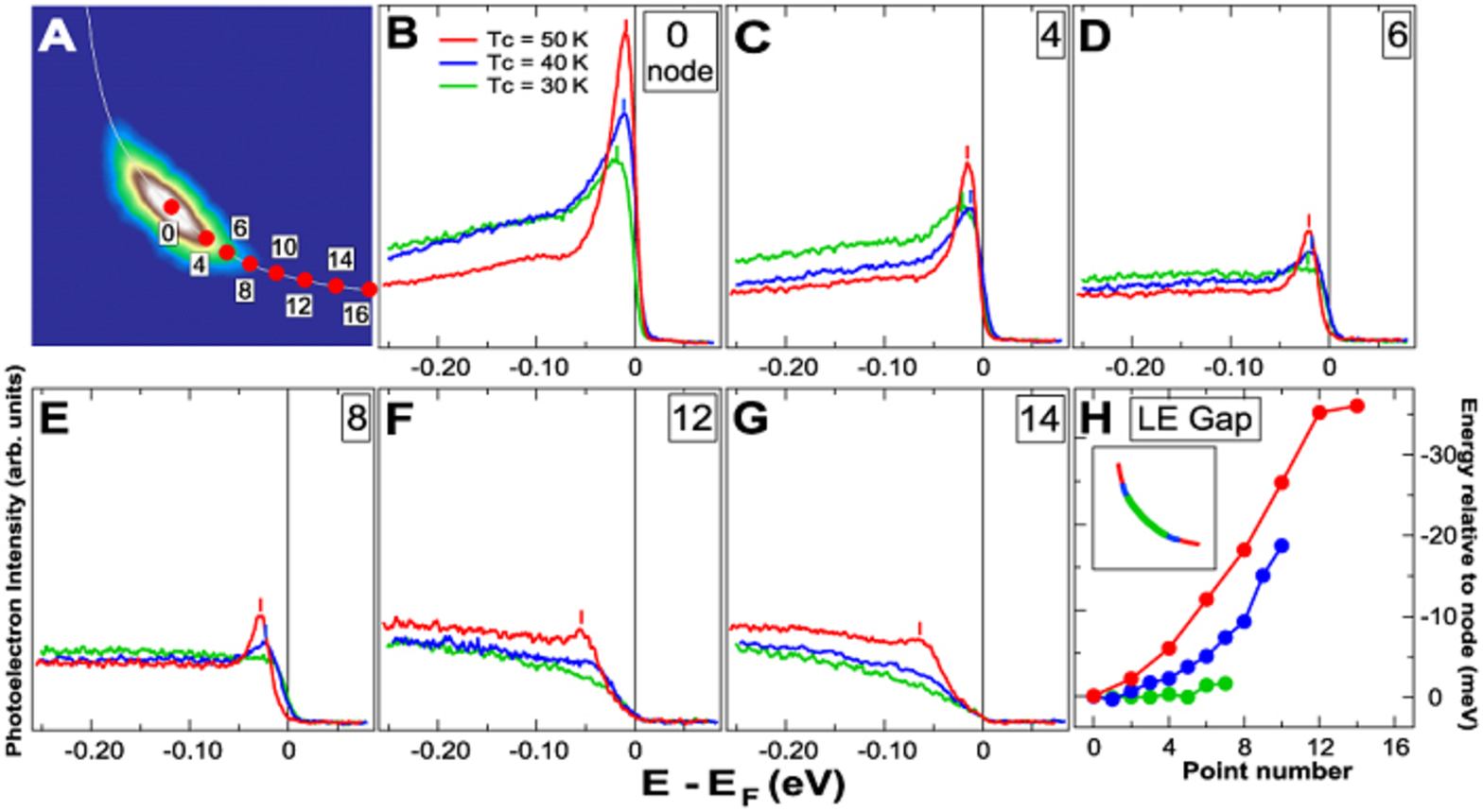} \caption{\textbf{Doping dependence of the ARPES spectra
and leading edge position along the Fermi surface.} \textbf{(A)}
Intensity plot of the spectral weight of the $T_c=30$ K sample,
which is integrated within $\pm10$ meV around the Fermi level
($E_{\rm{F}}$) and symmetrized with respect to the diagonal of the
zone. The intensity of diffraction replica due to supermodulation of
the crystal structure has been cut off by color scale. The red dots
labeled with numbers denote the Fermi crossing points $k_{\rm{F}}$
determined from the momentum distribution curves (MDC) at
$E_{\rm{F}}$. \textbf{(B to G)} The energy distribution curves (EDC)
along the Fermi surface (FS) of the $T_c$=30, 40, and 50 K samples.
The number shown at the upper right corner corresponds to the
$k_{\rm{F}}$ locations shown in (A). The vertical bars indicate the
peak position of the EDC. The $k_{\rm{F}}$ region of the FS where a
peak is visible in the EDCs is operationaly defined as the Fermi
arc. \textbf{(H)} Doping dependence of the leading edge position
relative to that of the nodal spectrum within the Fermi arc. The
inset illustrates that the Fermi arc elongates toward the antinodal
region of $k$-space with increasing doping. The data shown in this
paper were taken in the second Brillouin zone where an enhancement
of the spectra is observed (See Supporting Online Materials).}
\end{center}
\end{figure}

\begin{figure}[p]
\begin{center}
\includegraphics[width=170mm]{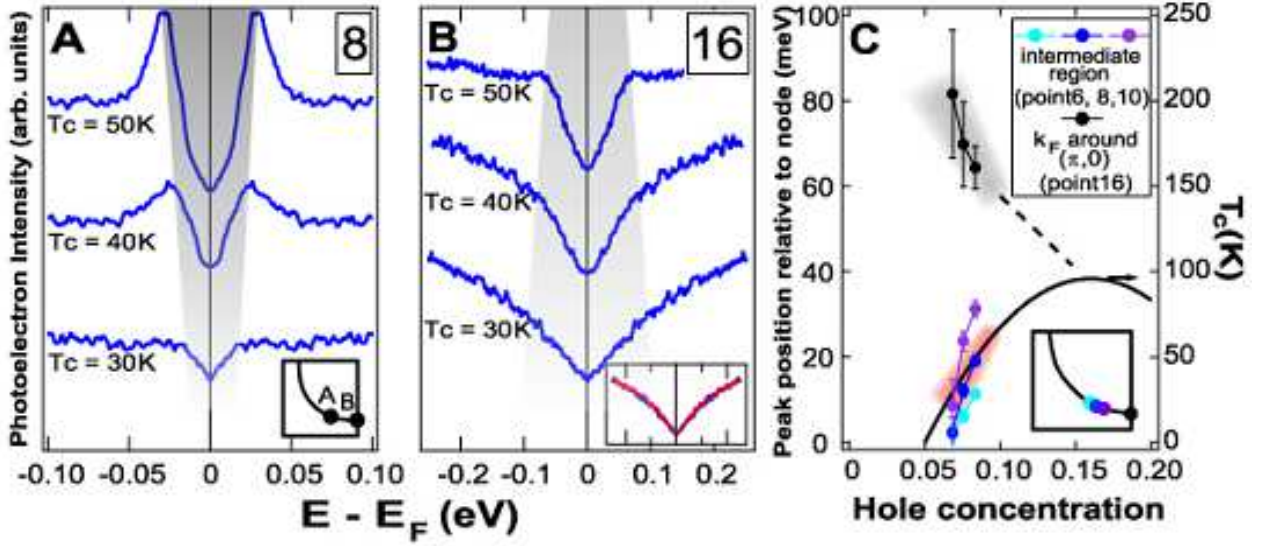}
\caption{\textbf{Doping dependence of the symmetrized spectra at the
intermediate region and the antinodal region.} The symmetrized EDCs
at \textbf{(A)} the intermediate region of the FS (point 8 in Fig.
1A) and \textbf{(B)} the antinodal region (point 16). Their
corresponding locations on the FS are shown in the inset of (A). The
shaded area denotes the region inside the gap determined by the peak
positions of the EDCs. For the antinodal spectra, the position of
the hump, which is determined from the second derivative of the
spectra, is used as the peak position. Inset of (B) shows
temperature dependence of the spectra of the $T_c=30$ K sample taken
at 10 K (blue line) and 50 K (red line) at the antinodal region.
\textbf{(C)} Doping dependence of the peak position around the
intermediate region (point 6, 8, and 10) and in the antinodal region
(point 16) shown in the inset together with $T_c$. The dashed line
indicates the pseudogap at the antinodal region reported by previous
ARPES studies on Bi2212 system \cite{Campuzano}. Here, the hole
concentration ($p$) was estimated by using an empirical relationship
$T_c=96(1-82.6(p-0.16)^2)$.}
\end{center}
\end{figure}

\begin{figure}[p]
\begin{center}
\includegraphics[width=170mm]{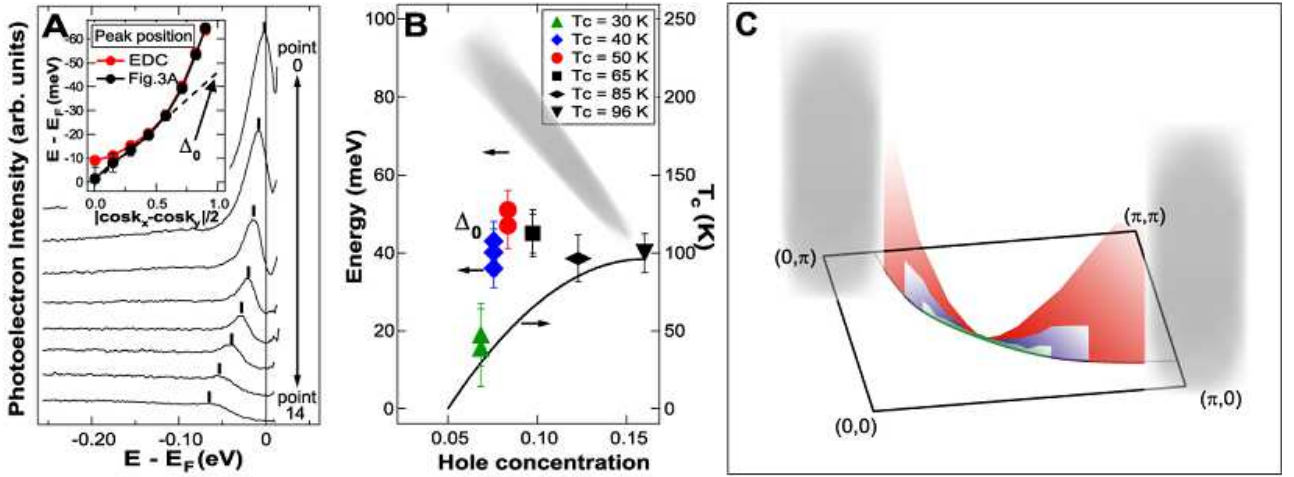}
\caption{\textbf{Determination of $\Delta_0$ and its doping
dependence.} \textbf{(A)} Spectra of the $T_c=50$ K sample along the
FS which were divided by the Fermi-Dirac (FD) function convoluted
with the experimental resolution. The vertical bars indicate the
peak position. The inset shows a comparison of the peak position in
EDCs plotted against the $d$-wave function
($\mid$cosk$_x$-cosk$_y$$\mid/$2) for the raw EDCs and the
FD-function divided EDCs. The dashed line illustrates an
extrapolation of the straight section around the nodal region of the
black curve, which leads to $\Delta_0$ at
$\mid$cosk$_x$-cosk$_y$$\mid/2=1$. \textbf{(B)} Doping dependence of
the $\Delta_0$. The gray shaded area is a guide to the eyes for the
doping dependence of the energy gap at the antinodal region from
data shown in Fig. 2C as well as existing published data
\cite{review:ARPES}. $\Delta_0$s for $T_c>60$ K were taken from our
ARPES studies of
Bi$_2$Sr$_2$Ca$_{0.92}$Y$_{0.08}$Cu$_2$O$_{8+\delta}$ samples (see
Supporting Online Material). \textbf{(C)} Doping dependence of the
leading edge position in the $k$-space. The same color assignment as
(B) is used for a easy comaprison.}
\end{center}
\end{figure}

\end{document}